\documentclass[a4paper,11pt]{article}
\usepackage{hepunits}
\usepackage{SIunits}
\usepackage{hepnames}
\usepackage{hepunits}
\usepackage{SIunits}
\usepackage{hepnames}

\pdfoutput=1 
             
\usepackage{jinstpub}

\title{\boldmath The upgrade of the ALICE Inner Tracking System}

\author{D.~Andreou}
\affiliation{CERN, 1211 Geneva 23, Switzerland and \\Nikhef, Amsterdam, The Netherlands}

\emailAdd{dimitra.andreou@cern.ch}

\abstract{The Inner Tracking System (ITS) of the ALICE experiment will be upgraded during the second long LHC shutdown in 2019--2020. The main goal of the ALICE ITS upgrade is to enable high precision measurements of low-momentum particles ($<1~\GeVoverc$) by acquiring a large sample of events, benefiting from the increase of the LHC instantaneous luminosity of Pb--Pb collisions to $\mathcal{L} = 6 \cdot 10^{27} \cm^{-2} \second^{-1} $ during Run 3. Working in this direction the ITS upgrade project is focusing on the increase of the readout rate, on the improvement of the impact parameter resolution, as well as on the improvement of the tracking efficiency and the position resolution. The major setup modification is the substitution of the current ITS with seven layers of silicon pixel detectors. The ALPIDE chip, a CMOS Monolithic Active Pixel Sensor (MAPS), was developed for this purpose and offers a spatial resolution of 5~$\micron$. The use of MAPS together with a stringent mechanical design allows for the reduction of the material budget down to 0.35\%~$X_0$ for the innermost layers and 1\%~$X_0$ for the outer layers.\\
The detector design was validated during the research and development period through a variety of tests ensuring the proper operation for the full lifetime inside ALICE. The production phase is close to completion with all the new assembled components undergoing different tests that aim to characterize the modules and staves and determine their qualification level. This contribution describes the detector design, the measurements performed during the research and development phase, as well as the production status.}


\collaboration[c]{on behalf of the ALICE collaboration}

\proceeding{9$^{\text{th}}$ International Workshop on Semiconductor Pixel Detectors for Particles and Imaging
\\10-14 December, 2018
\\Taipei, Taiwan}

\begin{document}
\maketitle
\flushbottom

\section{Introduction}
\label{sec:intro}
\begin{figure}[htbp]
\centering
\includegraphics[width=.5\textwidth,clip]{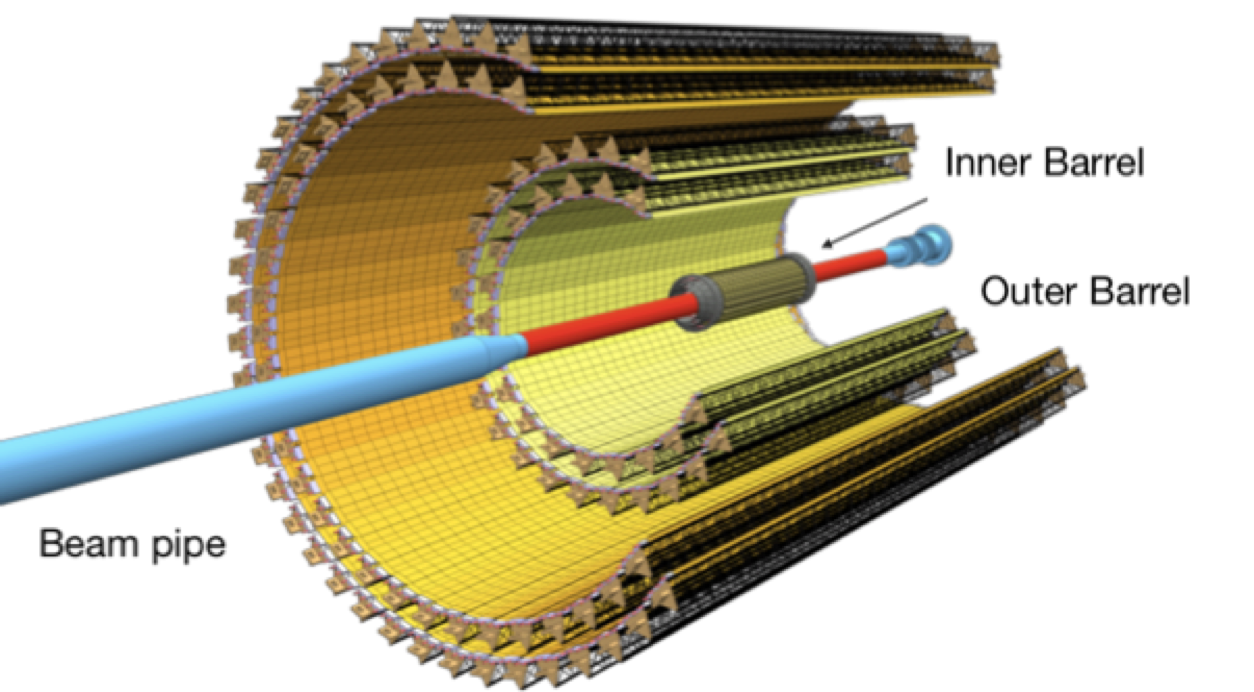}
\caption{\label{fig:layout} Layout of the upgraded ALICE ITS.}
\end{figure}

The ALICE  (A Large Ion Collider Experiment) experiment at CERN is designed to address the physics of strong interactions and in particular the properties of the Quark--Gluon Plasma (QGP). The innermost detector of ALICE, the Inner Tracking System (ITS), will be upgraded during the second long LHC shutdown in 2019-2020. The main physics goal of the ITS upgrade is to enable high precision measurements of rare probes over a broad $p_\mathrm{T}$ range especially in the region $<1~\GeVoverc$ \cite{a}.  \\
In order for ALICE to be able to accumulate the necessary data to achieve its physics goals, detector modifications are done that lead to the increase of the readout rate from the current 1~kHz  to 100~kHz (Pb--Pb collisions) and to the improvement of the tracking efficiency and the impact parameter resolution. The current ITS consists of six detector layers using three different technologies: Silicon Pixel (SPD), Silicon Drift (SDD) and Silicon Strip (SSD) detectors. These six layers will be replaced by seven layers of ALICE Silicon Pixel Detector (ALPIDE) chips which were developed for the upgrade of the ITS \cite{b}. The ALPIDE chip is a CMOS Monolithic Active Pixel Sensor (MAPS) combining the sensor and the readout electronics in one single device. The choice of MAPS allows the thinning of the detector and thus contributes to the reduction of the material budget. The pixel size is reduced as well from $50\times425$~$\micron^2$ (SPD) down to $29\times27$~$\micron^2$. Figure \ref{fig:layout} shows the overview of the upgraded ITS. The seven layers are divided into two barrels, the Inner Barrel (IB) which consists of the three innermost layers and the Outer Barrel (OB) with four outer layers. The OB is further divided into the two middle layers and the two outer layers which share the same design but have different length, 843~mm and 1475~mm respectively. The thickness of the ALPIDE chips in the IB is 50~$\micron$ which brings the material budget down to 0.35\%~$X_0$ whereas for the OB the silicon thickness is 100~$\micron$ contributing to the overall OB material budget of 1\%~$X_0$. 

The upgraded ITS will cover a 10~$\meter^2$ active surface with 12.5~billion pixels offering a radius coverage of 22--400 mm and a pseudorapidity coverage of $\mid{\eta}\mid$ < 1.22 for tracks from 90$\%$ of the luminous region. The detector will operate in a 0.5 T magnetic field and has to be able to withstand the radiation level of 2.7 Mrad TID and $1.7 \times 10^{13}$1MeVn$\mathrm{_{eq}~cm^{-2}}$ NIEL. All the above modifications will contribute to the improvement of the impact parameter resolution in the beam direction $z$ by a factor of six, from 240~$\micron$ to 40~$\micron$, and in the transverse plane $r\phi$ by a factor of three, from 120~$\micron$ to 40~$\micron$ at $p_\mathrm{T}$ $0.5~\GeVoverc$ . 

\section{The ALPIDE chip}
\begin{figure}[htbp]
\centering 
\includegraphics[width=.55\textwidth,clip]{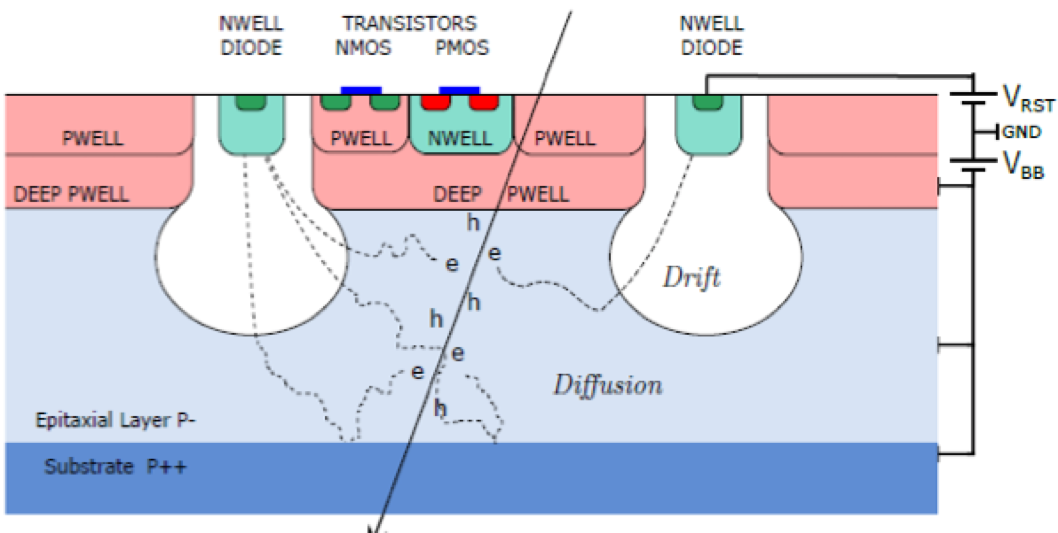}
\caption{\label{fig:tower}Cross section through a MAPS made in the TowerJazz 0.18~$\micron$ imaging process.}
\end{figure}

The ALPIDE chip is the building block of the ALICE tracker \cite{c}. The ALPIDE chips are produced by TowerJazz based on the 0.18~$\micron$ CMOS imaging process shown in Figure \ref{fig:tower}. A key feature of the process is the deep p-well that is shielding the n-well and thus allows implementing full CMOS in-pixel circuitry. The total ionizing dose tolerance requirement is achieved with small thickness gate oxides whereas the 25~$\micron$ p-type epitaxial layer with a resistivity >1 k$\Omega$ cm and the possibility of applying reversed bias voltage to the substrate increase the tolerance to the non-ionizing energy loss. All the above characteristics improve the charge collection and make the ALPIDE chip a unique choice for the ALICE tracker. Test beam and laboratory measurements on the ALPIDE chips have shown excellent detector performance after exposure in TID and NIEL radiation over a wide range of operational settings. The detection efficiency is higher than 99$\%$ and the spatial resolution is within the required 5~$\micron$ \cite{d}. 

The ALPIDE chip has dimensions $15\times30$~$\mm^2$ and hosts $512\times1024$ pixels.  Each pixel cell contains a sensing diode, a front-end amplifier and shaping stage, a discriminator, and a digital section. The pixels are arranged in double columns and read out by a priority encoder which sends the addresses of the pixels that recorded a hit to the chip periphery \cite{e}. Pixels with no hits are not read out making the readout procedure faster and reducing the power consumption which for every pixel is about 40~nW.
\section{Detector design}
\subsection{Outer and inner barrel modules and staves}
\begin{figure}[htbp]
\centering
\includegraphics[width=.49\textwidth,trim=0 -130 0 0,clip]{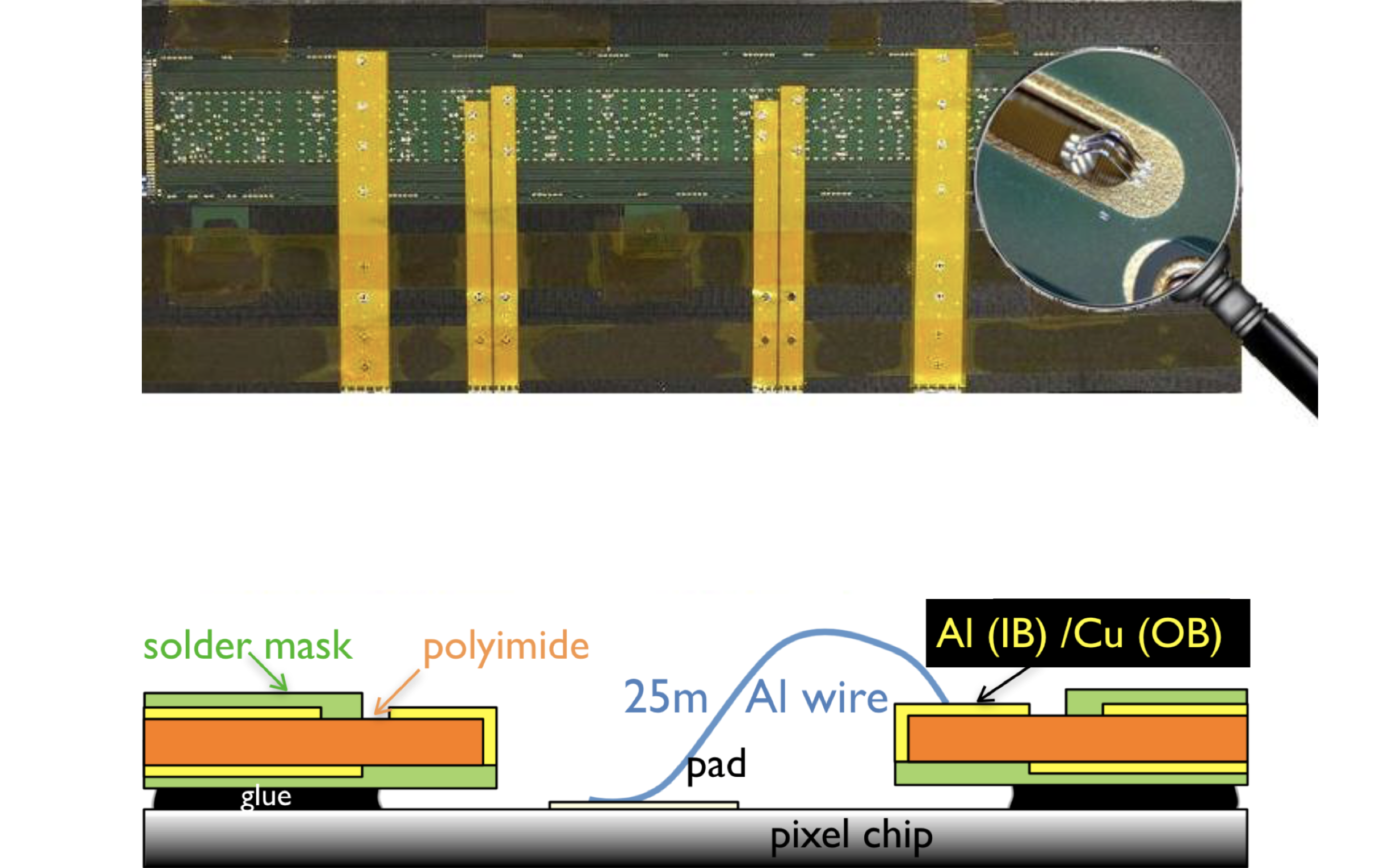}
\qquad
\includegraphics[width=.16\textwidth]{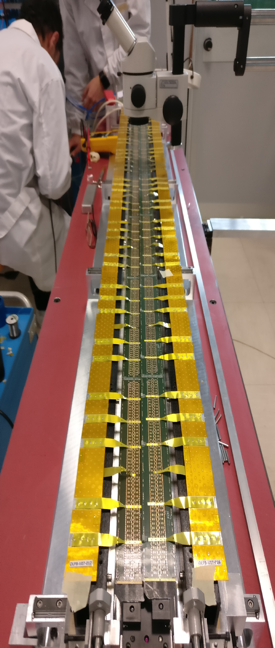}
\caption{\label{fig:OBstave} OB module with soldered cross cables (top left), FPC components (bottom left) and fully assembled OB stave consisting of two half-staves soldered to the power buses before they are folded on top (right).}
\end{figure}

The ALPIDE chips are assembled into modules which are mounted on staves. The detector design is similar for the IB and OB modules encountering differences that emerge from the different data rate in the two barrels. Starting from the OB, each OB module is made out of two rows of seven ALPIDE chips. The data rate in the OB is much lower and the number of chips much larger in comparison with the IB. For this reason the OB modules are read out in a master-slave topology, in which all data are transferred through two masters per 14-chip module. In the IB on the other hand, each chip has a dedicated high-speed data line. The 14 OB sensors are glued on a copper Flexible Printed Circuit (FPC) to which are electrically connected via wirebonds as shown in the left picture of Figure \ref{fig:OBstave}. The wirebonds connect the metallized vias in the FPC to pads distributed over the entire chip surface. The sensors and the FPC  form a Hybrid Integrated Circuit (HIC). The FPC distributes the clock, the configuration signals, the data readout, and the power. The power supply is done through aluminium cross cables that are soldered on the FPC. 

Concerning the OB staves, each stave consists of two half-staves. Each middle layer half-stave consists of four OB modules whereas the outer layer half-stave consists of seven OB modules. The base of the OB stave is a lightweight carbon mechanical space frame that hosts two cooling leakless structures, the cold plates. The power supply is distributed to all HICs in one half-stave via a power bus, which is connected to the HICs via a set of aluminium-based cross cables.

\begin{figure}[htbp]
\centering 
\includegraphics[width=.65\textwidth,clip]{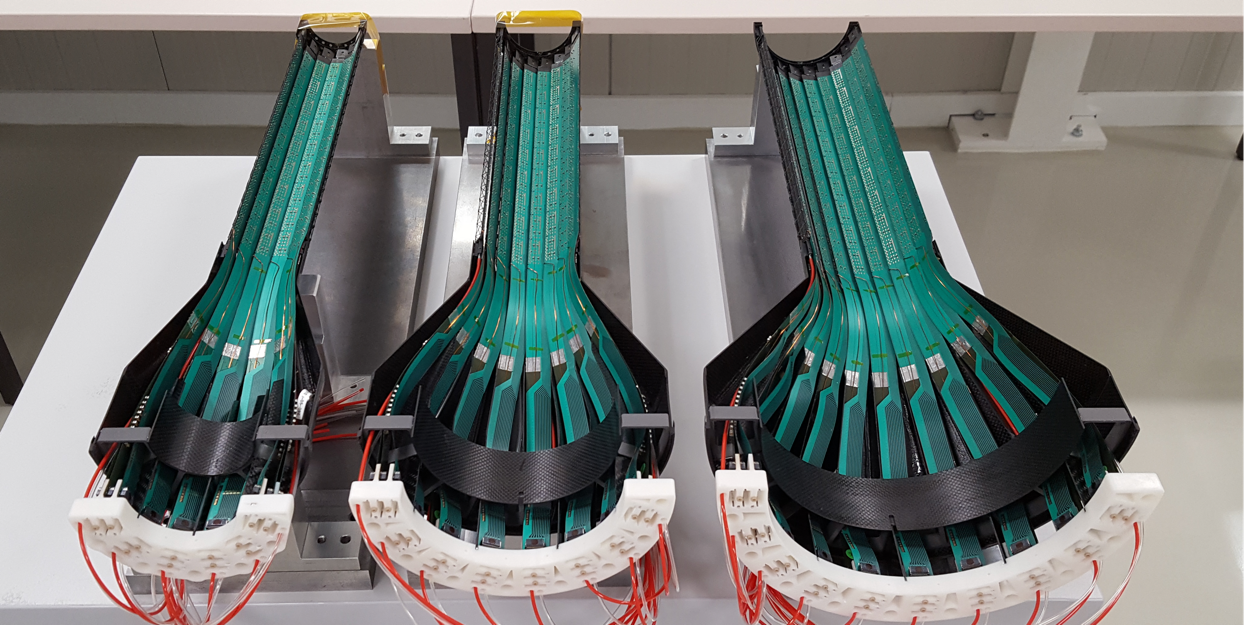}
\caption{\label{fig:iblayers} First assembled inner barrel half-layers.}
\end{figure}

In the IB case, each IB stave is made out of one IB module that consists of nine ALPIDE chips all operating in master mode. There is again the space frame and the cold plate and on top the sensors, bonded with the FPC. Each one of the ALPIDE chips on an IB module has an individual data line and all the chips are read out in parallel. The conducting layers of the FPC in the IB module are made out of aluminium. Figure \ref{fig:iblayers} shows IB staves assembled in the IB half-layers. Then the three half-layers are assembled all together and they make the half barrel. In the end the two half barrels are attached together and they make the full IB of the ITS.

\begin{figure}[htbp]
\centering 
\includegraphics[width=.7\textwidth,clip]{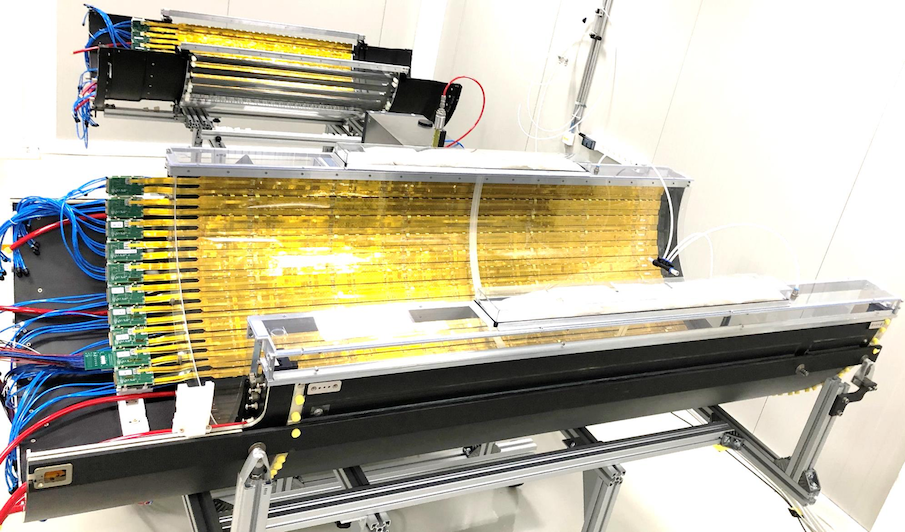}
\caption{\label{fig:mechanics} OB half-layer 6 and OB half layer 4 (on the back) with OB staves mounted.}
\end{figure}

The construction of the OB is done also by assembling two half outer barrels. The OB staves of each half-layer are inserted in a structural shell as shown in Figure \ref{fig:mechanics}. Each half barrel consists of four half layers. The fully assembled outer barrel is finally inserted in the cage barrel.
\subsection{Readout electronics}
\begin{figure}[htbp]
\centering 
\includegraphics[width=.35\textwidth,clip]{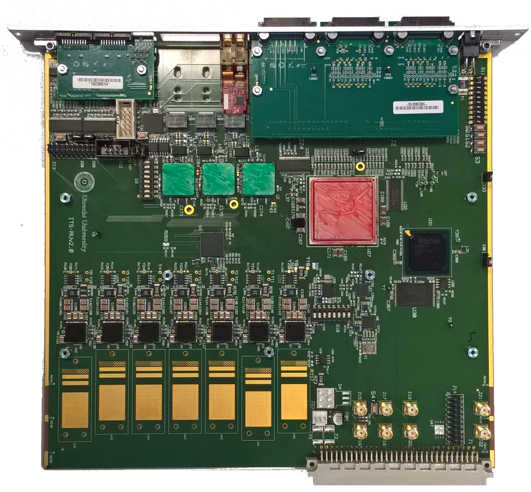}
\caption{\label{fig:ru} ALICE ITS readout unit.}
\end{figure}

The readout logic is fully implemented inside the ALPIDE chip which sends the digitized and zero-suppressed hit data to the off-detector electronics through 8~m long electrical differential links. In total 192 readout units (see Fig.  \ref{fig:ru}) will be used and each one of them, hosted into a 6U VME board, is connected to an entire IB or OB stave. The bandwidth for the data transmission is 1.2~Gb$/\second$ and 400~Mb$/\second$ for the IB and OB respectively. The readout units are then connected through optical links with the Common Readout Unit (CMU) and with the Central Trigger Processor (CTP).
\section{Production and qualification}
\begin{figure}[htbp]
\centering
\includegraphics[width=.3\textwidth]{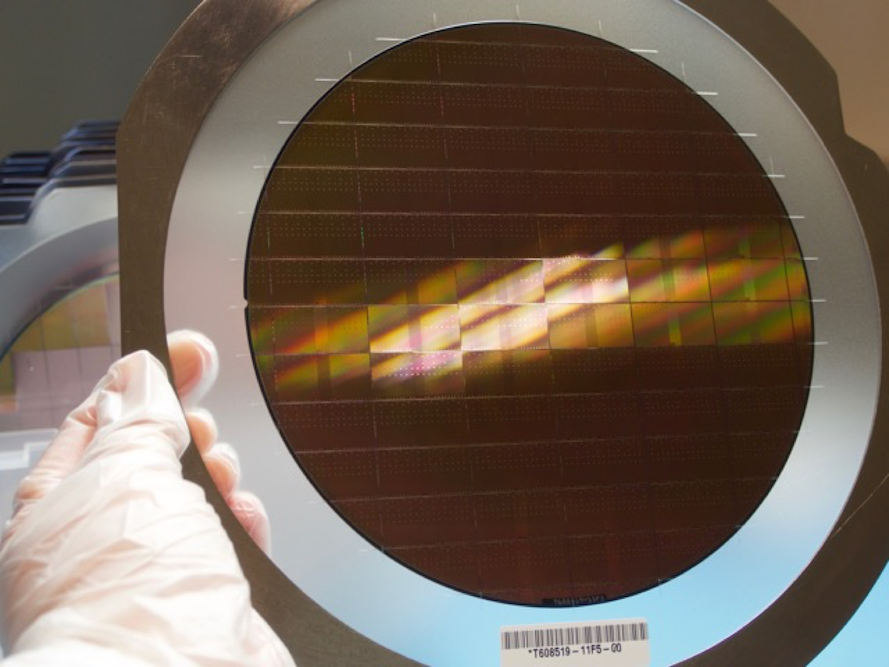}
\qquad
\includegraphics[width=.27\textwidth]{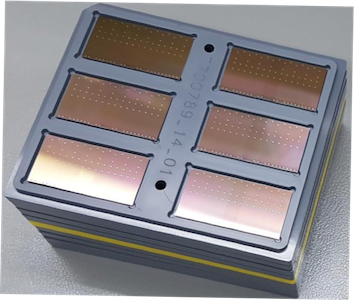}
\caption{\label{fig:wafer} Silicon wafers (left) and ALPIDE chips on trays (right). }
\end{figure}

The production of the ALPIDE chips starts with the CMOS wafer production by TowerJazz and continues with the thinning and dicing of the silicon wafers. The chips are picked up and placed inside trays (Fig.~\ref{fig:wafer}) and sent for the chip probe testing. After the chips are tested the 100 $\micron$ chips are distributed to five OB HIC assembly sites and the 50 $\micron$ ones to CERN for the IB HIC production.

\begin{figure}[htbp]
\centering
\includegraphics[width=.44\textwidth]{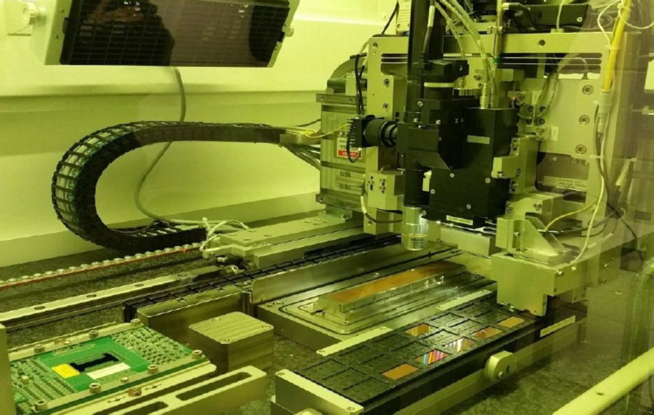}
\qquad
\includegraphics[width=.48\textwidth]{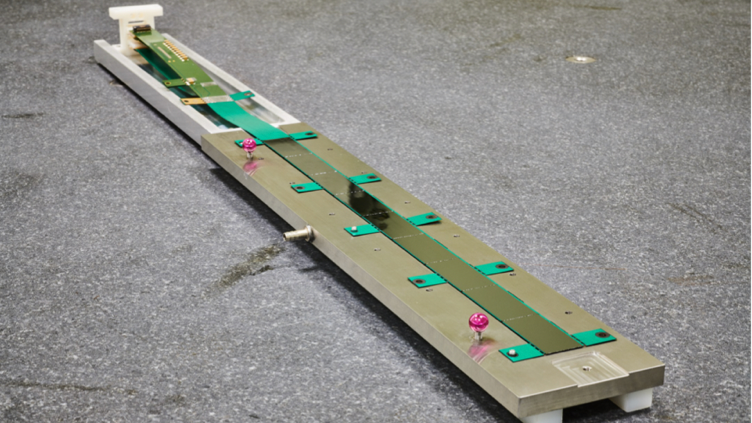}
\caption{\label{fig:ALICIA} Alignment and gluing of the chips on the FPC (left) and IB module (right). }
\end{figure}

\begin{figure}[htbp]
\centering 
\includegraphics[width=0.98\textwidth,clip]{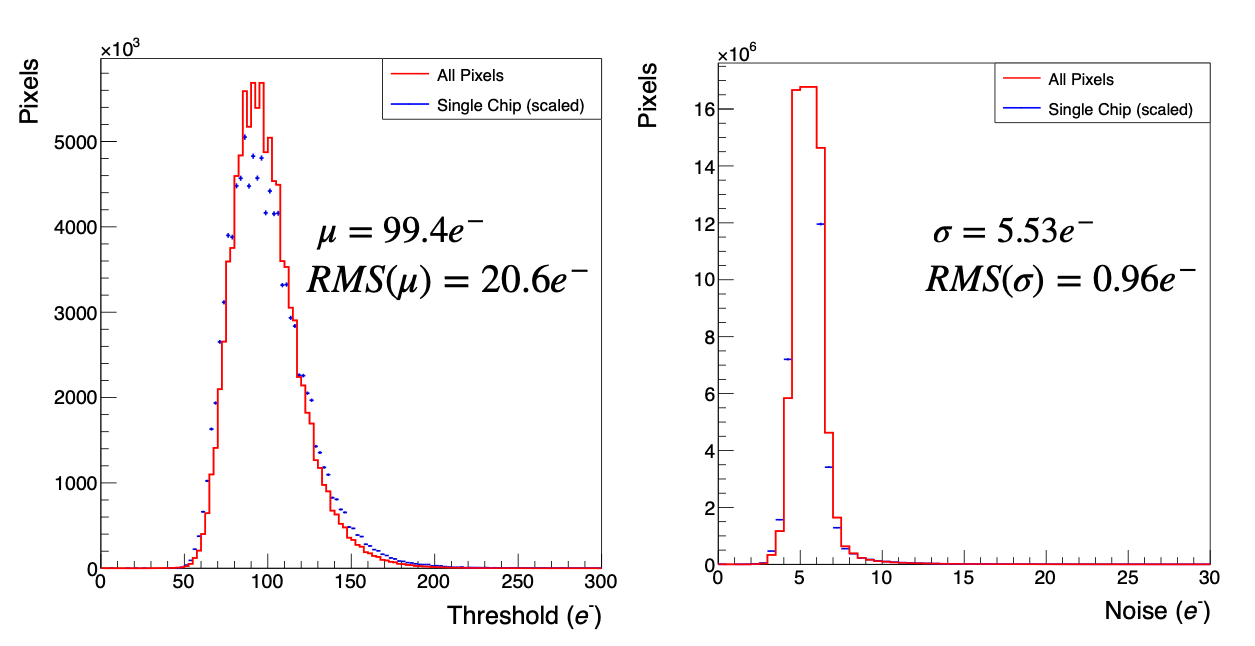}
\caption{\label{fig:thrnoisehistos} Threshold $\mu$ (left) and noise $\sigma$ (right) distribution of the 100~M pixels of an OB stave, after chip by chip threshold tuning. By adjusting the operational settings of the chips, a uniform chip response is achieved throughout the stave. The scaled single chip threshold distribution (blue) agrees with the threshold distribution of all pixels on the staves (red) indicating only pixel to pixel variations.}
\end{figure}

In the HIC production centers the chips are selected and the FPC verification takes place. Afterwards the chips are aligned and glued with high precision on the FPC  with the use of a custom made machine  (Fig. \ref{fig:ALICIA}). The electrical connection between the chips and the FPC is realized through wire bonding. The HIC production procedure is completed with the testing and the qualification of the HICs and their distribution to five stave assembly sites (OB case).
The modules are glued on the half-stave cold plates. The alignment of the modules is done using a set of dedicated alignment jigs guided by a coordinate measuring machine. The machine ensures that the modules are placed on the cold plate having a maximum distance of 100 $\micron$. After the curing of the glue, the HIC to HIC interconnections are established by soldering. The next step is the glueing of the two cold plates of the half-staves on the space frame. Finally the power buses are connected to the cross cables of the modules, folded and the stave is ready to be tested and qualified.\\
Detailed verification and testing is done throughout all the production steps. The assembled components undergo a number of electrical tests that are performed with the use of dedicated software. The functionality of the components is evaluated in different levels by testing the readout and analogue performance of the full matrix, the digital functionality of the periphery, and the electrical quality of the interface signals. The different scans are included in a user friendly Graphical User Interface (GUI) through which the modules and staves are tested and classified. In the end of the testing procedure all the information and the data are stored in a common database. Typical results of the threshold scan, which checks the analogue performance of the component under test, are shown in Figure \ref{fig:thrnoisehistos}.  
\newpage
Apart from the tests done for the qualification and classification of the components, more tests are performed during the production period for the validation and quality assurance of the assembly procedure. Peel tests are done on the modules to check the glue adhesion between the chips and the FPC, as well as pull tests to verify the wirebond strength.  \\
Multiple series of ageing tests were performed on IB and OB modules as well as on IB staves in order to validate the assembly procedure and ensure the component quality for a ten year operation period in the ALICE cavern conditions.  The ageing tests included thermal cycles in the temperature range from $10^\circ$ to $50^\circ$ C, which correspond to the lower temperature the detector will experience in case the detector stops operating but the cooling is still active and to the maximum temperature during normal operation in case of a cooling failure respectively. The components were also tested and exposed to a high humidity environment corresponding to 60$\%$ Relative Humidity (RH) in order to accelerate the ageing. The ageing tests were divided into periods that corresponded to single years inside the ALICE cavern. Each one of the years consisted of a number of thermal cycles and a  humidity test. Electrical tests and visual inspection were performed after each one-year equivalent. The results of the ageing tests verified the assembly procedure and validated the use of the appropriate components. The electrical tests demonstrated no difference in the digital and analogue performance of the modules  and staves and the visual inspection reported no difference in the material quality. 
\begin{figure}[htbp]
\centering
\includegraphics[width=.45\textwidth]{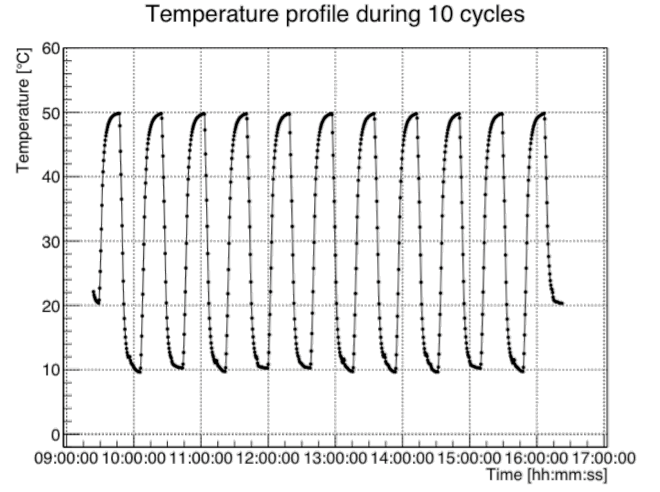}
\qquad
\includegraphics[width=.45\textwidth]{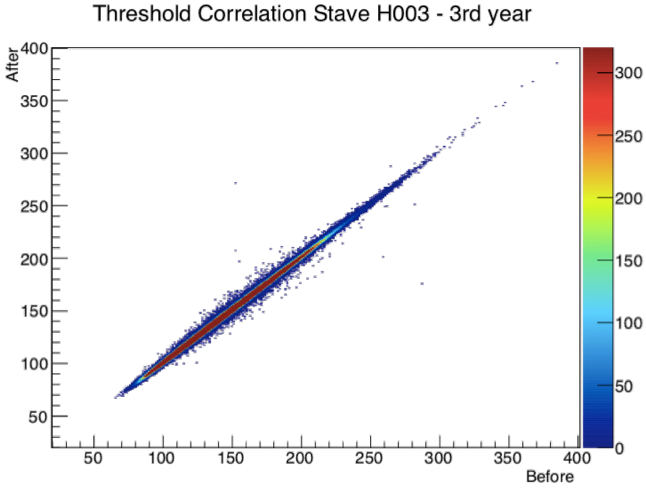}
\caption{\label{fig:ageing} Temperature distribution over ten cycles (left) and threshold correlation of an IB stave after three years of ageing tests (right). }
\end{figure}
\section{Summary and outlook}
The upgraded ITS, a 12.5 Gpixel detector, will be installed in the ALICE cavern in 2020. The production period of the ITS upgrade will be completed the summer of 2019 with the production of the readout electronics and the construction of the outer barrel. The production of the ALPIDE chips, the mechanical support structures and all the IB layers has been completed. The full detector assembly will take place during the 2019. The implementation of the CMOS technology in the ALICE tracker along with a low material budget detector design will improve the impact parameter resolution, the tracking efficiency and the readout rate capabilities opening a new era of measurements for the ALICE experiment.

\end{document}